\documentclass[a4paper,pre,twocolumn,superscriptaddress,showkeys,floatfix]{revtex4-1}
\usepackage{graphicx}
\usepackage{amsmath}
\usepackage{amssymb}
\usepackage{centernot}

\begin{document}

\title{Damage spreading in the random cluster model}

\date{\today} 

\author{P. H. Lundow} 
\email{per.hakan.lundow@math.umu.se} 


\affiliation{Department of mathematics and mathematical statistics,
  Ume\aa{} University, SE-901 87 Ume\aa, Sweden}

\begin{abstract}
  We investigate the damage spreading effect in the Fortuin-Kasteleyn
  random cluster model for 2- and 3-dimensional grids with periodic
  boundary. For 2D the damage function has a global maximum at
  $p=\sqrt{q}/(1+\sqrt{q})$ for all $q>0$ and also local maxima at
  $p=1/2$ and $p=q/(1+q)$ for $q\lesssim 0.75$. For 3D we observe a
  local maximum at $p=q/(1+q)$ for $q\lesssim 0.46$ and a global
  maximum at $p=1/2$ for $q\lesssim 4.5$. The chaotic phase of the
  model's $(p,q)$-parameter space is where the coupling time is of
  exponential order and we locate points on its boundary.  For
  3-dimensional grids the lower bound of this phase may be equal to
  the corresponding critical point of the $q$-state Potts model for
  $q\ge 3$.
\end{abstract}


\maketitle

\section{Introduction}
The study of damage spreading has its early origin in statistical
systems where binary vectors are iteratively mapped by randomly chosen
Boolean functions. If two independently chosen vectors then are
exposed to the same sequence of functions, how does their Manhattan
distance behave with time and when do the vectors become equal? (see
for example Ref.~\cite{derrida:86} and references therein). In the
1980s similar questions were asked for Ising spin
systems~\cite{stanley:87}. Here a replica of an equilibrated
2-dimensional spin system is made with one spin changed, giving it an
initial damage. Updating the two systems with the Glauber rule while
using the same sequence of random numbers to guide the updating, the
difference between the two systems, the damage $D_t$, then evolves
with time $t$. For low temperatures $T<T_c$ the two systems quickly
become identical ($D_t=0$ at $t=\tau$) but for high temperatures
$T>T_c$ the damage spreads through the replica, reaching a positive
fraction of the spins ($D_t>0$).


The damage spreading phenomenon has, since then, mainly been studied
for Ising spin glass systems but also in many other
contexts~\cite{derrida:87, arcangelis:89, campbell:91, campbell:93,
  lundow:12, bernard:10, rubio:13, rubio:16}.  Let us describe the
phenomenon in the spin-glass case in slightly more detail.  Here two
independent and randomly chosen $\pm 1$-spin vectors of length $N$,
having the same interaction strengths $J_{ij}$, are exposed to the
same sequence of random numbers for updating the spins, this time with
heat-bath dynamics.

Now the situation becomes the opposite of that just described for the
Ising case~\cite{stanley:87,derrida:87}: above a certain temperature
$T_{D}$, in the ordered (``frozen'' or ``stable'') phase the two spin
vectors become identical after just a few time steps (say,
$\tau=\mathcal{O}(N)$) and then stays identical, i.e., the two systems
have coupled. Below this temperature, in the chaotic (or disordered)
phase, the coupling time $\tau$ instead diverges exponentially with
$N$ and the two systems never seem to meet.  In this phase, the
distance $D_t$ fluctuates around some temperature-dependent mean value
$D=\langle D_t\rangle$. In short, at $T_{D}$ the coupling time $\tau$
changes (with decreasing $T$) from staying bounded, through
logarithmic and polynomial growth, to exponential growth, see Fig.~3
of Ref.~\cite{bernard:10}.

The opposite behaviour, compared to the Ising case, is due to the
different dynamics: heat-bath versus Glauber
updating~\cite{stanley:87}. Recall that with heat-bath dynamics a spin
is {\it set} to its new value with some probability, whereas for
Glauber dynamics a spin is instead {\it flipped} with some
probability.  Note also that for spin glasses, $T_{D}$ is not the same
as the critical temperature $T_c$, as in the 2D Ising case. For
example, for 2D systems where $J_{ij}=\pm 1$ with equal probability,
we have $T_{D}\approx 1.69$~\cite{lundow:12} even though $T_c=0$.

Here we instead investigate damage spreading in the setting of the
random-cluster model where two parameters $p$ and $q$ control the
distribution of states~\cite{FK:72,grimmett:06}. With this model it is
the state of edges (on/off) rather than the state of spins ($\pm 1$)
that is updated.  The underlying graphs will be 2- and 3-dimensional
grids of linear order $L$. Recall that the random cluster model
contains the percolation model ($q=1$) and the $q$-state Potts model
(integer $q\ge 2$) which in turn generalises the Ising model without
an external field.

For smaller $L$ we obtain an support set $S_L$ of the damage function
$D(p,q)$ by simply updating a system pair for a long time, depending
on $L$, to see if $D\to 0$ in that time. The chaotic phase $S$ is
obtained from estimating the growth rate of the coupling time for many
choices of parameters $p,q$. Effectively the chaotic phase is a limit
of the support sets, $S_L\to S$. However, determining $S_L$ for larger
$L$ can be computationally demanding near its boundary, but the small
$L$ tells us where to look for the boundary of $S$. Thus it can
actually be easier to estimate the boundary of $S$ by simply measuring
growth rates of the coupling time.

We have computed the damage function and estimated the support set for
2D and 3D systems by scanning a grid of $p,q$-parameters and for a
range of $L$. Based on this we have approximated the boundary of the
chaotic phase. When $q<1$ for 2D grids, this boundary has a smooth,
almost triangular shape, but is considerably more complex for 3D
grids.  We also guess simple formulas describing the boundary. For 3D
grids with $q\ge 3$ the lower bound of the chaotic phase seems to be
closely related to the critical point of the $q$-state Potts model.

\section{Definitions and details}
The underlying graph $G=(V,E)$ is the $d$-dimensional grid graph of
order $L$ with periodic boundary conditions thus having $N=L^d$
vertices and $m=dL^d$ edges. The partition function of the
Fortuin-Kasteleyn (FK) model~\cite{FK:72}, for parameters $0<p<1$ and
$q>0$ is
\begin{equation}\label{zfun}
  Z(p,q) = \sum_{A\subseteq E} p^\epsilon (1-p)^{m - \epsilon} q^\kappa
\end{equation}
where $\epsilon=\epsilon(A)=|A|$ is the number of edges in $A$ and
$\kappa=\kappa(A)$ is the number of (connected) components, or
clusters, in the spanning subgraph $G(V,A)$. For convenience, we will
overload the notation and write $\epsilon(p,q) =
\langle\epsilon\rangle/m$ and $\kappa(p,q)=\langle\kappa\rangle/N$ for
the normalised mean values under the distribution of states.  The
parameter $p$ controls the edge-density, in fact $\epsilon(p,q)$ is
non-decreasing in $p$ for fixed $q>0$~\cite{grimmett:06}.  Parameter
$q$ controls the bias towards clusters; $q>1$ means we prefer many
clusters, $q<1$ means we prefer less of them, and $q=1$ gives a random
subset of the edges $E$ without any such bias.

We will describe the process of updating $A$ in some detail.  The
state of an edge $e\in E$ is either $e\in A$ or $e\notin A$.  We wish
to update this state by choosing at random whether it should be
reversed (flipped). Reversing the state of $e=\{u,v\}$ changes
$\epsilon$ by $\Delta\epsilon=\pm 1$. However, $\Delta\kappa$ depends
both on the state of $e$ and whether there is a $uv$-path in $A$ {\it
  not} using $e$, here denoted $u\leftrightsquigarrow v$. The four
possible cases are then
\begin{subequations}\label{ecases}
  \begin{eqnarray}
    1.\quad&e\in A \land u\leftrightsquigarrow v:
    \quad\Delta\epsilon=-1,&\quad\Delta\kappa=0 \\
    2.\quad&e\in A \land u\centernot\leftrightsquigarrow v:
    \quad\Delta\epsilon=-1,&\quad\Delta\kappa=1 \\
    3.\quad&e\notin A \land u\leftrightsquigarrow v:
    \quad\Delta\epsilon=+1,&\quad\Delta\kappa=0\\
    4.\quad&e\notin A \land u\centernot\leftrightsquigarrow v:
    \quad\Delta\epsilon=+1,&\quad\Delta\kappa=-1
  \end{eqnarray}
\end{subequations}

An edge $e\in E$ selected at random will then belong to one of these
cases and we denote the probability for these cases $P_1, P_2, P_3,
P_4$.  Alternatively, the underlying graph being edge transitive, as
the present grid graphs, we might also say that for a fixed edge, the
probability that this edge belongs to case $i$ is $P_i$.  Note, by the
way, that $P_1+P_2$ is the probability that $e\in A$ so that
$\epsilon(p,q)=P_1 + P_2$.

As usual, briefly put, our update rule should maintain so-called
detailed balance between the current set $A$ and the next set $A'$,
differing only at the edge $e$,
\begin{equation}\label{balance}
  \Pr(A)\Pr(A\to A') = \Pr(A')\Pr(A'\to A)
\end{equation}
The probability ratio between the states is
\begin{equation}\label{pratio}
  x = \frac{Pr(A')}{\Pr(A)} = \left(\frac{p}{1-p}\right)^{\Delta\epsilon} q^{\Delta\kappa}
\end{equation}
With $F(x)=\Pr(A\to A')$ we have $F(1/x)=\Pr(A'\to A)$ and thus we
want a function satisfying $0\le F(x)\le 1$ and $F(x) = x F(1/x)$.
Choosing a random number $U$ uniformly distributed over the interval
$\lbrack 0,1)$, the update rule is now simply: reverse the state of
edge $e$ if $U\le F(x)$.

We have mainly used the traditional $F(x)=\min(1,x)$ of Metropolis
fame, but we will also briefly use $F(x)=\tfrac{x}{1+x}$ (heat-bath)
for comparison.  Functions like $F(x)=\tfrac{x+x^2+\ldots
  +x^k}{1+x+x^2+\ldots+x^k}$ work equally well but do not seem to be
used in the literature. See Ref.~\cite{grimmett:06} for a more
complete description of Metropolis/heat-bath in the random cluster
context and Ref.~\cite{binder:73} for more on update rules.

The process starts at time step $t=0$ with a random subset $A\subseteq
E$ where $\Pr(e\in A)=\Pr(e\notin A)=1/2$. A second set $B\subseteq E$
is initially chosen as the complement of $A$, i.e., $B=E\setminus A$.
The damage can now be defined as $D=\tfrac{1}{m}|(A\cup B)\setminus
(A\cap B)|$, or,
\begin{equation}\label{damage}
  D=\frac{1}{m}\sum_{e\in E} |\mathcal{I}(e\in A)-\mathcal{I}(e\in B)|
\end{equation}
where $\mathcal{I}(s)$ is an indicator function returning $1$ if $s$
is true and $0$ otherwise.  At $t=0$ the damage is thus always $D=1$.
Both sets are now exposed to exactly the same sequence of random
numbers and the edges in both sets are updated according to the
protocol described above. A time step consists of performing a sweep
where we update $m$ randomly selected edges from $E$. The same
sequence of edges are then candidates for addition/deletion at both
sets and they are tried with the same random numbers.

If we at some time step $t$ receive $D_t=0$, the coupling time $\tau$
has been reached and the damage will stay zero at all future time
steps. The damage will eventually go to zero but the coupling time
divides the $p,q$-parameter space into a chaotic phase and an ordered
phase.  In the ordered phase, the coupling time is of at most
polynomial order, $\tau=\mathcal O(L^c)$ for some finite $c$, thus
potentially $D_t\to 0$ comparatively fast. In the chaotic phase the
coupling time is instead super-polynomial, typically exponential.
Except for small $L$ we will in practice never obtain $D_t=0$ in this
phase. Instead, the damage stabilizes after some equilibration, around
a value $D(p,q)=\langle D_t\rangle$.

When moving along a line in the $p,q$-parameter space, from the
chaotic to the ordered phase, we typically find $\tau = \mathcal
O(L^c)$ at a single point. We will take a simple view on the
transition between the two phases, going from exponential, via
polynomial at a point, to subpolynomial, say logarithmic or bounded
$\tau$. However, we do not wish to rule out that this transition may
be more complex than this.

For a given $L$ and after some practical number of equilibration
steps, we can treat $D(p,q)$ as, effectively, a function with support
set $S_L=\{(p,q):D(p,q)>0\}$. The chaotic phase will then be the set
of $(p,q)$-points where $\tau$ is of superpolynomial order, i.e.,
$\tau=\omega(L^c)$ for all constants $c$, using little-omega notation.
At least for small $L$, where $\tau$ is still not too large, it is
practical to view $S$ as a limit set of $S_L$.  We will denote by
$S(q)$ the set for fixed $q$. Denote by $\underbar{S}(q)$ and
$\bar{S}(q)$ the lower and upper bounds for fixed $q$.

For $d=2$ we define
\begin{equation}\label{pc2d}
  p_c(q)=\frac{\sqrt{q}}{1+\sqrt{q}}
\end{equation}
which for $q\ge 1$ is known to be the critical point of the random
cluster model~\cite{beffara:12,grimmett:06}, though we will use this
notation also for $q<1$.  For $d=3$ no general formula for $p_c(q)$ is
known but estimates exists of course for specific integer $q$-values
in the context of Ising- or Potts-model estimates.  Recall that the
translation between the temperaure parameter $\beta$ of the $q$-state
Potts model and $p$ of the random cluster model is through the simple
relation $p=1-e^{-\beta}$, see for example
Refs.~\cite{blote:05,grimmett:06}.

For $d=3$ this gives $p_c(2)=0.358091335(6)$ (via Ising model
$\beta_c$-estimate~\cite{ferrenberg:18}), $p_c(3)=0.42330(5)$ (Potts
model estimates, see for example Refs.~\cite{lundow:09,janke:97}),
$p_c(4)=0.4666(3)$ (see for example
Refs.~\cite{chatelain:05,babaev:15}).  For $q=5$ the phase transition
becomes first order and $\beta_c$ is harder to estimate. We are
unfortunately not aware of any published estimates of $\beta_c$ for
$q\ge 5$.

In the present work we have collected random cluster data at a large
number of $p,q$-points for $L=6,7,\ldots,32$ when $d=2$.  and for
$L=3,4,\ldots,20$ when $d=3$. For both dimensions we also have data
for some larger $L$ but at much fewer $p,q$-points.  We also have
gathered coupling data at many points around the border of $S$,
sometimes for larger $L$.  All data were collected on a MacBook Pro
laptop computer and a PowerMac desktop computer, both with an 8-thread
processor, running programs written in Fortran.  All code can of course
be obtained from the author.

\section{Trajectories and coupling time}
\begin{figure}
  \includegraphics[width=0.483\textwidth]{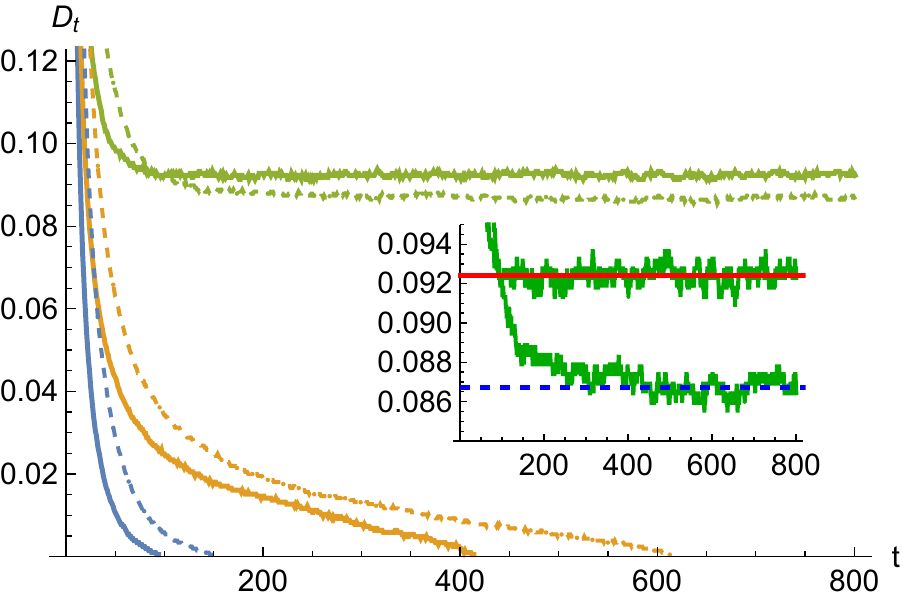}
  \caption{\label{fig0}(Colour on-line) $d=2$, $L=32$,
    $q=1/2$. Median of the damage $D_t$ taken over $1000$ trajectories
    for $p=0.27$ (blue, lower pair), $p=0.28$ (orange, middle pair)
    and $p=0.29$ (green, upper pair).  Each pair of curves are both
    Metropolis (solid curves) and heat-bath (dashed curves)
    updating. Inset shows a zoom-in of the $p=0.29$-curves.}
\end{figure}

In Fig.~\ref{fig0} we show the median damage trajectories for a
$32\times 32$-grid for three different $p$-values with $q=1/2$. With
Metropolis updating for $p=0.27$ the damage reaches 0 at $\tau=94(2)$,
for $p=0.28$ at $\tau=415(10)$ while for $p=0.29$ the damage does not
reach zero within this time frame. Using instead heat-bath updating
the coupling times change to $\tau=150(2)$ and $\tau=620(20)$
respectively. Thus we expect larger coupling times with heat-bath
updating. As the inset of Fig.~\ref{fig0} shows, for $p=0.29$ using
the Metropolis rule the damage stabilises very quickly to $D\approx
0.0924$ after roughly $200$ sweeps while for the heat-bath rule the
equilibration takes considerably longer, ca $600$ sweeps and then ends
up at a lower value, $D\approx 0.0868$. These examples suggest that
the Metropolis rule leads to a faster convergence both in the ordered
and the chaotic phase. It should here be mentioned that at
$p_c(1/2)\approx 0.4142$ we see no difference between the respective
$D$, when $t\gtrsim 50$, though the heat-bath curve is still slower to
equilibrate.

\begin{figure}
  \includegraphics[width=0.483\textwidth]{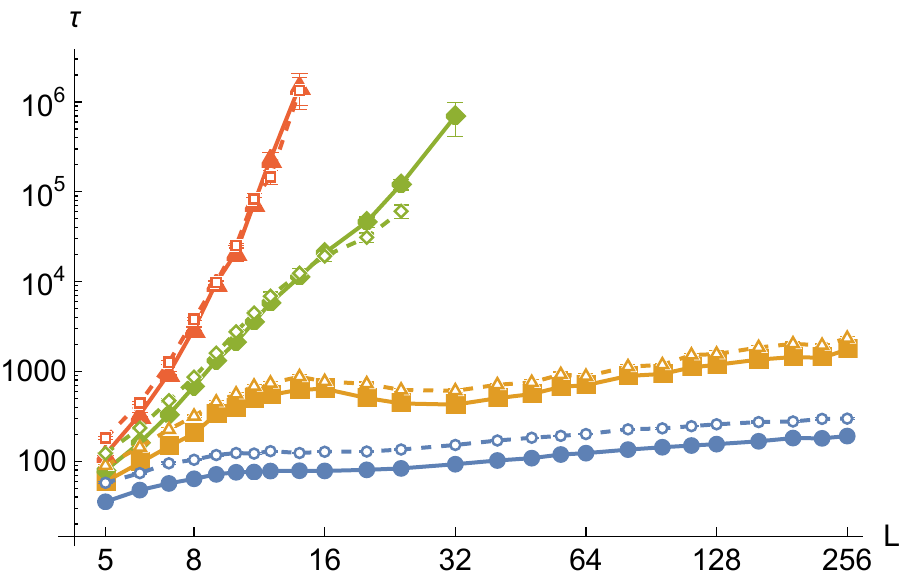}
  \caption{\label{fig1}(Colour on-line) $d=2$, $q=1/2$. Median
    coupling time versus $L$ for (upwards in plot) $p=0.27$ (blue
    pair), $p=0.28$ (orange pair), $p=0.285$ (green pair and dashed
    line) and $p=0.29$ (red pair).  Each pair of curves contains
    Metropolis (solid curves, filled markers) and heat-bath (dashed
    curves, open markers) updating.}
\end{figure}

Let us now instead look at the coupling time $\tau$ plotted versus
$L$.  In Fig.~\ref{fig1} we show the median $\tau$ taken over
many trajectories (error bars from bootstrap method) for a range of
$L$ and for the same $p,q$-values as in Fig.~\ref{fig0}, for both
heat-bath and Metropolis updating.  Clearly the coupling time differ
roughly by a factor for these $p$-values and we will henceforth only
focus on the Metropolis updating.

In principle one should be able to locate a $p$-value on the boundary
of $S(q)$ as a straight line in a log-log plot of $\tau$ versus $L$.
This is often true, but in this case we see rather strong finite-size
effects of $\tau$, which complicates matters.  At this point we can
only say for certain that $p=0.29$ is in the chaotic phase since the
coupling time clearly grows exponentially and is well-fitted by the
estimate $\tau = 0.59(3)\exp(1.06(1) L)$ for the Metropolis rule.

At $p=0.27$ the coupling time grows logarithmically and is well-fitted
over $L\ge 24$ by $\tau = -68(2) + 46(1) \log L$.  For $p=0.28$ the
small-$L$ effects are too strong to obtain a useful fit. Thus, in the
presence of such strong finite-size effects we receive a rough
estimate, $\underbar S(1/2) = 0.28(1)$.

\begin{figure}
  \includegraphics[width=0.483\textwidth]{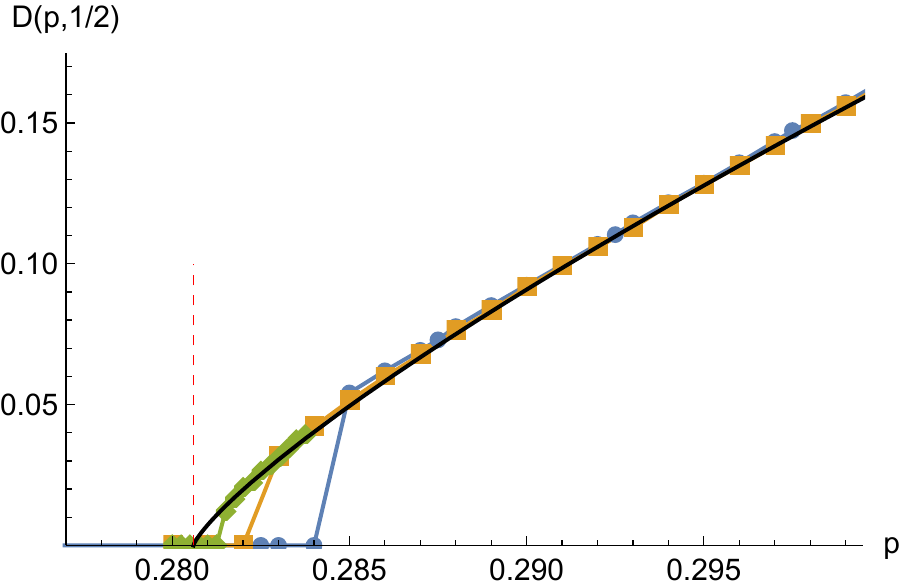}
  \caption{\label{fig2}(Colour on-line) $d=2$, $q=1/2$. Damage
    $D(p,1/2)$ versus $p$ for $L=32$ (blue points), 64 (orange
    squares) and 128 (green diamonds). Black curve is
    $D=3.8(p-0.2806)^{0.80}$. Vertical line at $p=0.2806$. }
\end{figure}

Let us instead look at the damage function $D(p,1/2)$ near $p=0.28$ as
shown in Fig.~\ref{fig2}. Damage curves for a range of $L$ agree on a
common limit curve and the curve stay positive closer to the boundary
for larger $L$.  From the data points in the plot we estimate the
limit curve $D = 3.80(p-0.2806)^{0.80}$, giving an excellent fit.
This approach gives the estimate $\underbar S(1/2)=0.2806(2)$. One can
thus resort to this approach when $\tau$ has strong finite-size
effects if high accuracy is important. Usually, however, we will
settle for the accuracy in the first approach.

\begin{figure}
  \includegraphics[width=0.483\textwidth]{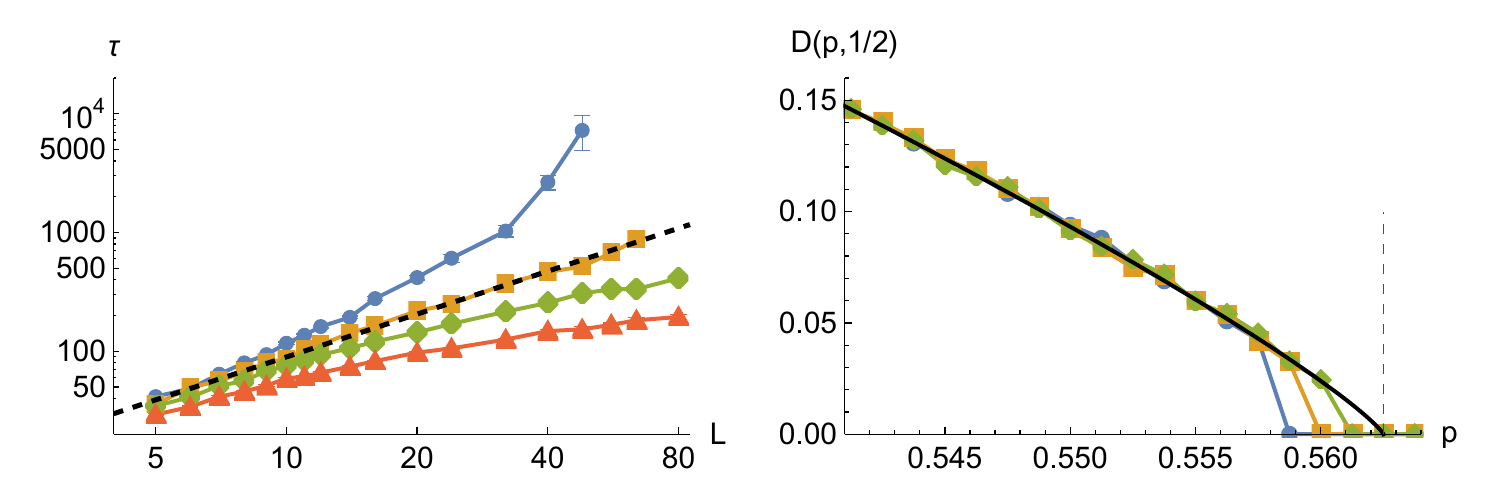}
  \caption{\label{fig3}(Colour on-line) $d=2$, $q=1/2$. Left: log-log
    plot of median coupling time $\tau$ versus $L$ for (downwards in
    plot) $p=0.56$ (blue points), $p=0.5625$ (orange squares),
    $p=0.565$ (green diamonds) and $p=0.57$ (red triangles). Black
    line (dashed) through $p=0.5625$ corresponds to $\tau =
    5.6L^{1.2}$. Right: Damage $D(p,1/2)$ versus $p$ for $L=32$ (blue
    points), 40 (orange squares) and 64 (green diamonds). Black curve
    is $D=3.78(0.5625-q)^{0.845}$.  Vertical red line at $p=0.5625$.}
\end{figure}

We will also show another example, see Fig.~\ref{fig3}, near the
upper bound $\bar S(1/2)$ where the finite-size effects are
negligible. For $p=0.5625$ the $\tau$-data are acceptably fitted by
a line in the log-log plot, giving $\tau = 5.6(3)L^{1.20(2)}$.  For
$p=0.56$ we already see an exponential behaviour, estimated to $\tau =
50(6)\exp(0.10(1)L)$. At $p=0.565$ and $p=0.57$ a distinctly
subpolynomial scaling is seen, but it does not scale logarithmically.
For $p=0.565$ we instead estimate it to be superlogarithmic, $\tau =
90(30) - 95(20) \log L +38(4)\log^2 L$.  From this we obtain the
estimate $\bar S(1/2)=0.5625(25)$.

In the right panel of Fig.~\ref{fig3}, the damage $D(p,1/2)$ is
shown. Again, from standard log-log fits, we find the simple formula
$D = 3.78(0.5625 - p)^{0.845}$ which fits these curves very
well. Unfortunately, due to slow equilibration, it is much more
difficult to obtain good $D$-data at this end of $S(1/2)$ and it would
require larger systems to improve significantly on the accuracy in our
estimate of $\bar S(1/2)$. Hence, locating the boundary of $S$ by
searching for polynomial order $\tau$ is actually practical in this
case. This will in fact be the main approach used to locate the
boundary of $S$.  The accuracy of the estimates are indicated by error
bars but often the error is smaller than the points in the
figures. Having established our approach we can now show the chaotic
phase for $d=2,3$ in the coming sections.

\section{The chaotic phase, $d=2$, $0<q<1$}
\begin{figure}
  \includegraphics[width=0.483\textwidth]{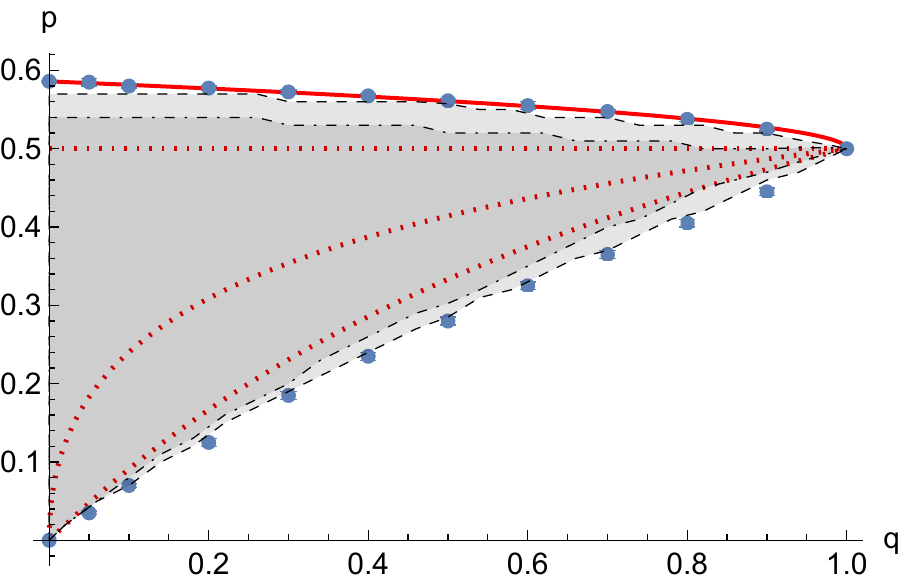}
  \caption{\label{fig5}(Colour on-line) $d=2$. Support $S_L$ for $L=8$
    (dot-dashed) and $L=32$ (dashed).  The curves (red, dotted) are
    $p=0.5+0.086\sqrt{1-q}$, $p=1/2$, $p=p_c(q)$ and $p=p_c(q^2)$
    (downwards). Points indicate boundary of $S$.}
\end{figure}

A rough estimate of $S$ can be obtained by equilibrating $D$ for as
many sweeps as possible, over a fine grid of $p,q$-points, for some
$L$. We show these for $L=8$ and $32$ in Fig.~\ref{fig5}.  The points
indicate estimated boundary points of $S$, including also the points
$(p,q)=(0,0)$ and $(1,1/2)$.  The upper bound is closely approximated
by the simple formula $\bar S(q)=0.50 + 0.086(2)\sqrt{1-q}$, but the
lower bound $\underbar S(q)$ has a more complicated behaviour for
which we have no suggested formula.

\begin{figure}
  \includegraphics[width=0.483\textwidth]{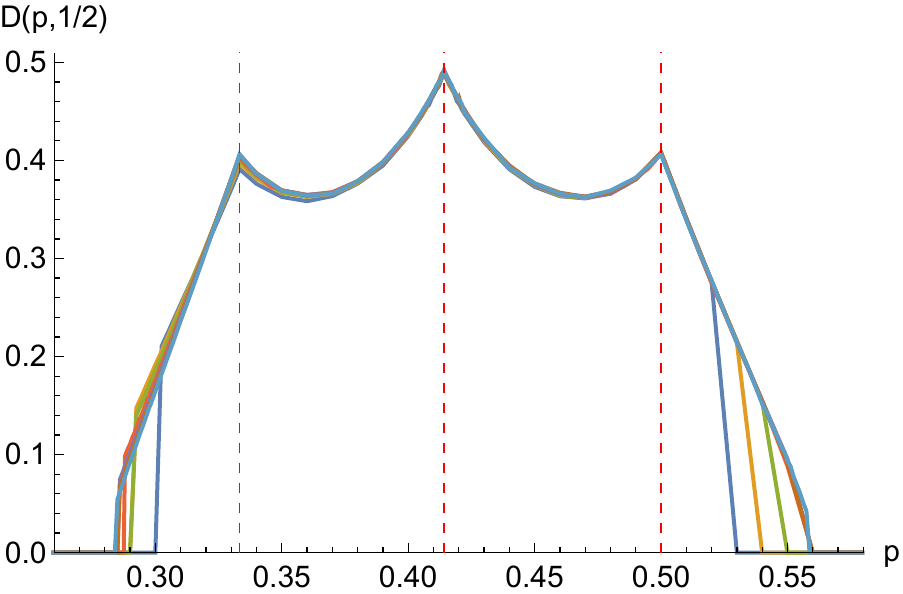}
  \caption{\label{fig6}(Colour on-line) $d=2$, $q=1/2$. Damage
    $D_L(p,1/2)$ versus $p$ for $L=8, 10, 12, 16, 20, 24,
    32$. The vertical lines (red, dashed) are at $p=1/3$,
    $p=\sqrt{2}-1$ and $p=1/2$. See Figs.~\ref{fig2} and \ref{fig3}.}
\end{figure}

The dotted lines in Fig.~\ref{fig5} indicate the location of three
local maxima of $D(p,q)$ for fixed $q$, namely: $p=1/2$, $p=p_c(q)$
and $p=p_c(q^2)=q/(1+q)$, see Eq.~\eqref{pc2d}.  This is demonstrated
in Fig.~\ref{fig6} for $q=1/2$ where the peaks are located exactly, as
far as we can tell, at $p=1/2$, $p=p_c(1/2) = \sqrt{2}-1$ and
$p=p_c(1/4)=1/3$. This picture is quite representative for $q\lesssim
0.75$, while for $q\gtrsim 0.75$ we see instead a triangle-shaped
function with only one cusp-like maximum at $p=p_c(q)$.

\begin{figure}
  \includegraphics[width=0.483\textwidth]{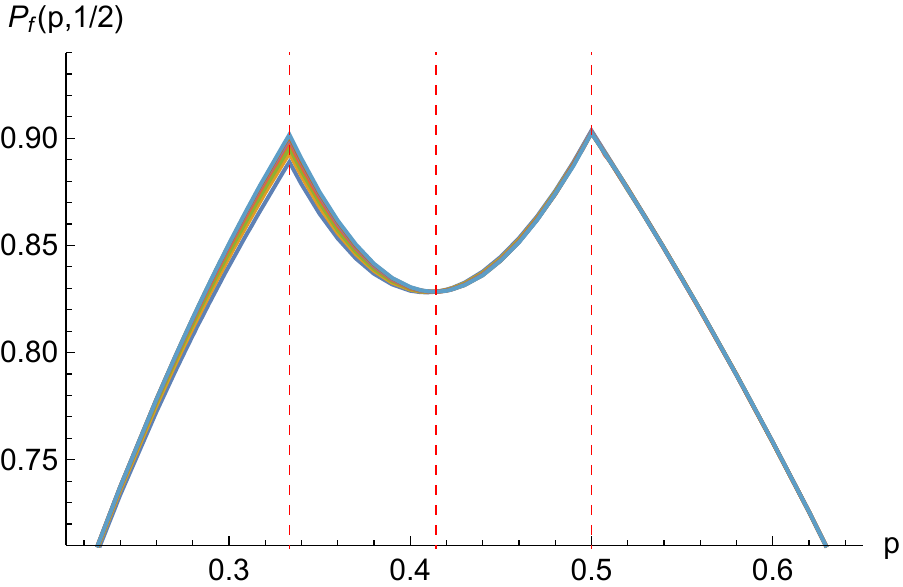}
  \caption{\label{fig7}(Colour on-line) $d=2$, $q=1/2$. Edge-flip
    probability $P_f$ versus $p$ for $L=8,10,12,16,20,24$. $P_f$
    increasing with $L$ around $p=1/3$.  The vertical lines (red,
    dashed) are at $p=1/3$, $p=\sqrt{2}-1$ and $p=1/2$.}
\end{figure}

The global maximum of $D$ at $p=p_c(q)$ supports that Eq.~\eqref{pc2d}
indeed is a critical $p$-value also for $q<1$ as well, seemingly
without finite-size corrections. It is far from clear why $p=1/2$ and
$p=q/(1+q)$ also would give local maxima.  Note that both $p=1/2$ and
$p=q/(1+q)$ gives cusp-like singularities in the edge-flip probability
$P_f$. This is not surprising since $P_f$ is a linear combination of
$P_i$ and $F(x)$ with $x$ depending on the cases of
Eq.~\eqref{ecases}. With $F(x)=\min(1,x)$ (Metropolis) both of these
$p$-values give this effect on $P_f$, see Fig.~\ref{fig7}.  We find
this effect in the damage function remarkable but can only note the
apparent relationship with the edge-flip probability.

The damage function differs very little when using heat-bath updating
but the edge-flip probability is instead quite smooth without any
cusps as in Fig.~\ref{fig6}.

\section{The chaotic phase for $d=2$, $q>1$}
\begin{figure}
  \includegraphics[width=0.483\textwidth]{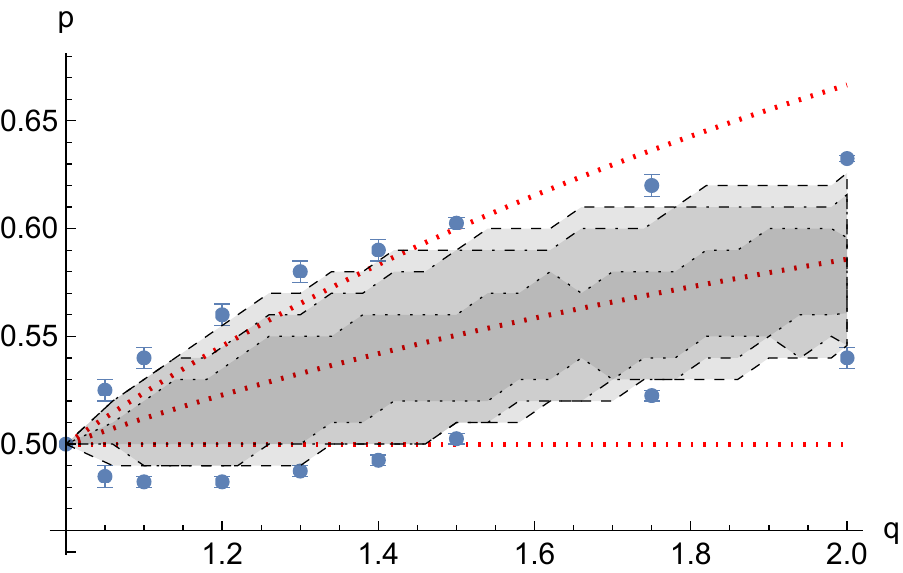}
  \caption{\label{fig8}(Colour on-line) $d=2$, $1<q\le 2$. Support
    $S_L$ for $L=8$ (innermost region, dotted boundary), $L=16$
    (dot-dashed boundary) and $L=32$ (outer region, dashed boundary).
    The curves (red, dotted) are $p=p_c(q^2)$, $p=p_c(q)$ and $p=1/2$
    (downwards). Points indicate boundary of $S$.}
\end{figure}

\begin{figure}
  \includegraphics[width=0.483\textwidth]{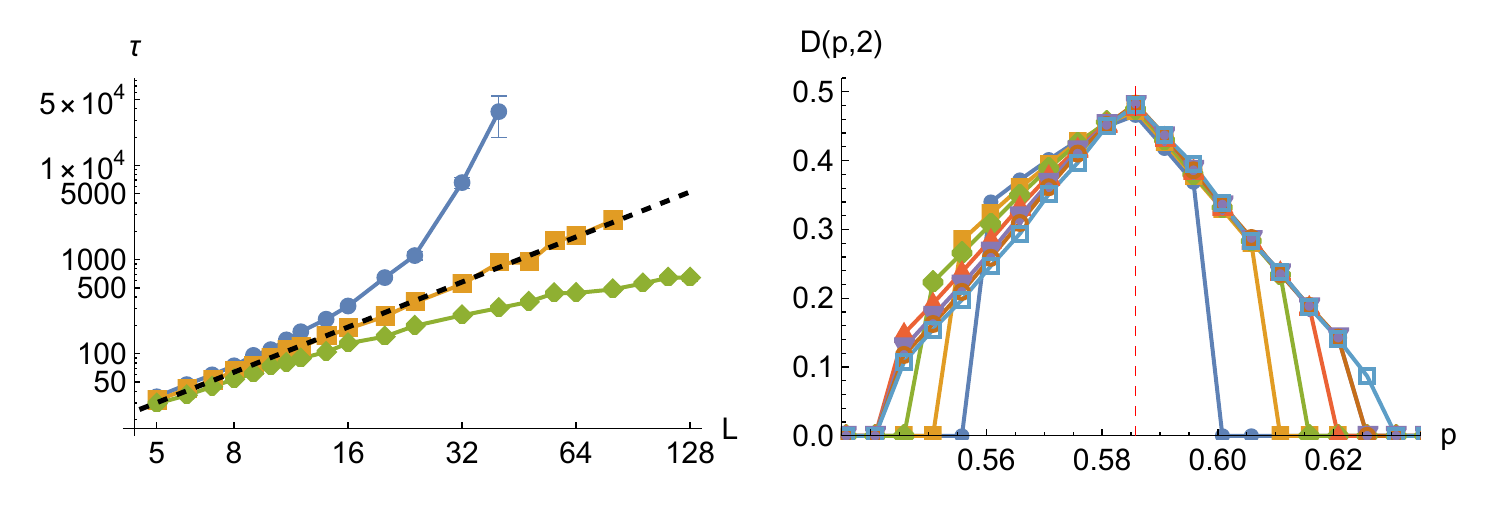}
  \caption{\label{fig9}(Colour on-line) $d=2$, $q=2$. Left: log-log
    plot of the median coupling time $\tau$ versus $L$ for (downwards
    in plot) $p=0.63$ (blue points), $p=0.6325$ (orange squares),
    $p=0.635$ (green diamonds). The dashed line corresponds to $\tau =
    2.3 L^{1.59}$, see text. Right: damage $D_L(p,2)$ versus $p$ for
    $L=10,12,16,20,24,32$. Larger systems has $D>0$ farther out from
    the maximum. Vertical red line located at the maximum,
    $p=p_c(2)=0.585786$.}
\end{figure}

For $q>1$ the chaotic phase surrounds $p_c(q)$ in a narrow band as
shown in Fig.~\ref{fig8} for $1<q\le 2$. Clearly the finite-size $S_L$
is growing outwards, approaching $S$. Estimating the boundary of $S$
proceeds as for $q<1$. We note that the lower bound of $S$ is
sometimes difficult to locate due to substantial finite-size effects
in $\tau$. However, almost no finite-size effects are seen for the
upper bound of $S$. An example is shown in the left panel of
Fig.~\ref{fig9} where we show the behaviour of $\tau$ near $\bar
S(2)$.

At the boundary, estimated to $\bar S(2)=0.6325(25)$, we see $\tau$ of
polynomial order, $\tau = 2.3(2)L^{1.59(3)}$. At $p=0.63$ we have
clear exponential growth in $\tau$ and subpolynomial $\tau$ at
$p=0.635$ (but probably not logarithmic). We have also estimated the
lower bound to $\underbar S(2)= 0.535(5)$, having a larger error bar
due to the previosly mentioned small-$L$ effects.

The right panel of Fig.~\ref{fig9} shows the damage function $D$ at
$q=2$ for a range of $L$. There are stronger finite-size effects away
from the maximum. Just as for $q<1$, the damage maximum seems to be
located at exactly $p_c(q)=\sqrt{2}/(1+\sqrt{2})$, for all $L$ and for
all $1<q\le 6$ that we have checked.

\section{The chaotic phase for $d=3$, $0<q<1$}
\begin{figure}
  \includegraphics[width=0.483\textwidth]{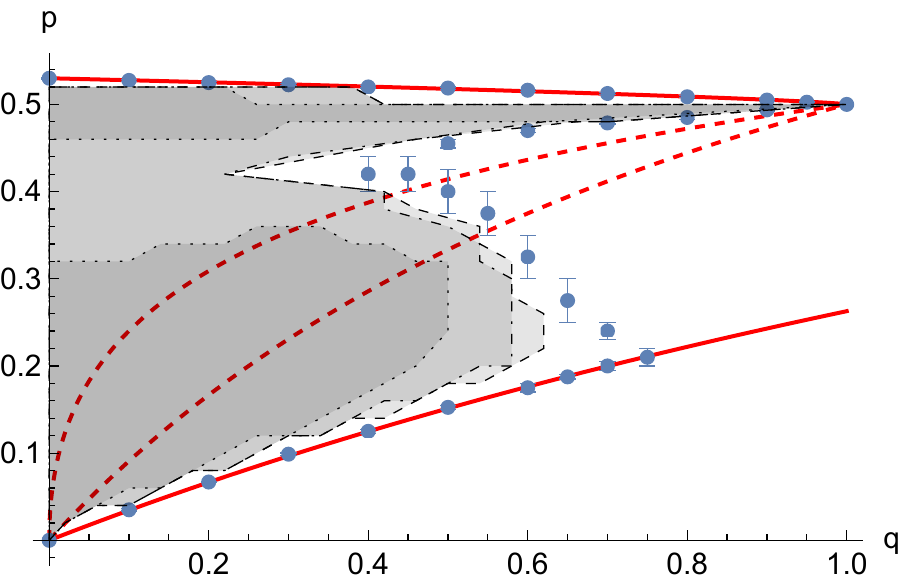}
  \caption{\label{fig10}(Colour on-line) $d=3$. Support $S_L$ for
    $L=12$ (dashed boundary), $L=10$ (dot-dashed boundary) and $L=6$
    (dark grey disconnected region, dotted boundary). Points
    indicate boundary of $S$. The solid curves (red) are
    $p=0.50+0.03(1-q)^{0.75}$ (upper boundary $\bar S(q)$) and
    $p=q/(2.805+q)$ (lower boundary $\underbar S(q)$).  The dotted
    curves (red) are $p=\sqrt{q}/(1+\sqrt{q})$ and $p=q/(1+q)$.}
\end{figure}

For 3-dimensional systems the chaotic phase has a quite different
shape from that of $d=2$. In Fig.~\ref{fig10} we show the support sets
(gray) for $L=6,10,12$ after several thousand sweeps.  For $L\le 9$
these sets are disconnected, splitting into two parts along $p\approx
0.42$. The points in the figure (with error bars) show our attempt at
pinpointing the boundary of the chaotic phase.  It turns out that a
good approximation for the lower bound is the simple $\underbar S(q) =
q/(2.805(1)+q)$, for $q\lesssim 0.75$, while the upper bound is
excellently fitted by $\bar S(q)=0.50+0.0300(7) (1-q)^{0.75}$ for
$q<1$. Both these curves are shown in Fig.~\ref{fig10}.

A complicating feature of this region is the wedge-shaped lower right
boundary which stretches up to roughly $(p,q)\approx
(0.21,0.75)$. Then it climbs upwards to the left in the figure to
about $(p,q)\approx (0.42,0.40)$ before finally taking a more
spike-shaped form.  The error bars are often rather large due to quite
significant finite-size effects in $\tau$. Determining the correct
shape here will require a more ambitious computational investment.  In
most other cases around the boundary, the finite-size effects are
negligible, thus giving very small error bars.

\begin{figure}
  \includegraphics[width=0.483\textwidth]{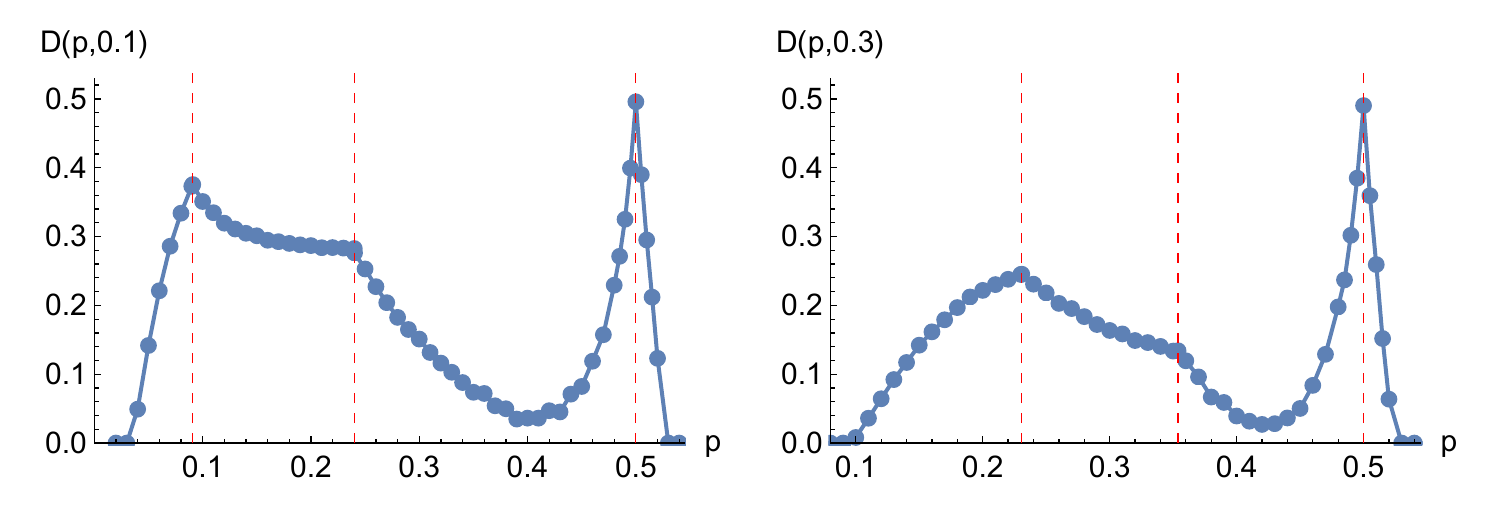}
  \caption{\label{fig11}(Colour on-line) $d=3$, $L=16$. Distance
    $D(p,0.10)$ (left) and $D(p,0.30)$ (right) versus $p$. The
    vertical lines (red, dashed) are at $p=q/(1+q)$,
    $p=\sqrt{q}/(1+\sqrt{q})$ and $p=1/2$ (see text).}
\end{figure}

The damage function $D(p,q)$ has a considerably less regular behaviour
for $d=3$ than for $d=2$. We show an example for $q=0.1$ and $0.3$ in
Fig.~\ref{fig11}.  For fixed $q$ there is a cusp-like maximum at
$p=1/2$ for all $0<q<1$. There is also a cusp-like maximum at
$p=q/(1+q)$, though it is no longer a maximum for $q\gtrsim
0.46$. More peculiar is that there is a cusp also at
$p=\sqrt{q}/(1+\sqrt{q})$ of Eq.~\eqref{pc2d}, but again this vanishes
for $q\gtrsim 0.46$. We have no explanation for why a critical point
for 2D would occur in the 3D-case. For $0.46\lesssim q \lesssim 0.66$
a new local maximum appears to the left of $p=q/(1+q)$, but we do not
have a conjecture for this point.

\section{The chaotic phase for $d=3$, $q>1$}
\begin{figure}
  \includegraphics[width=0.483\textwidth]{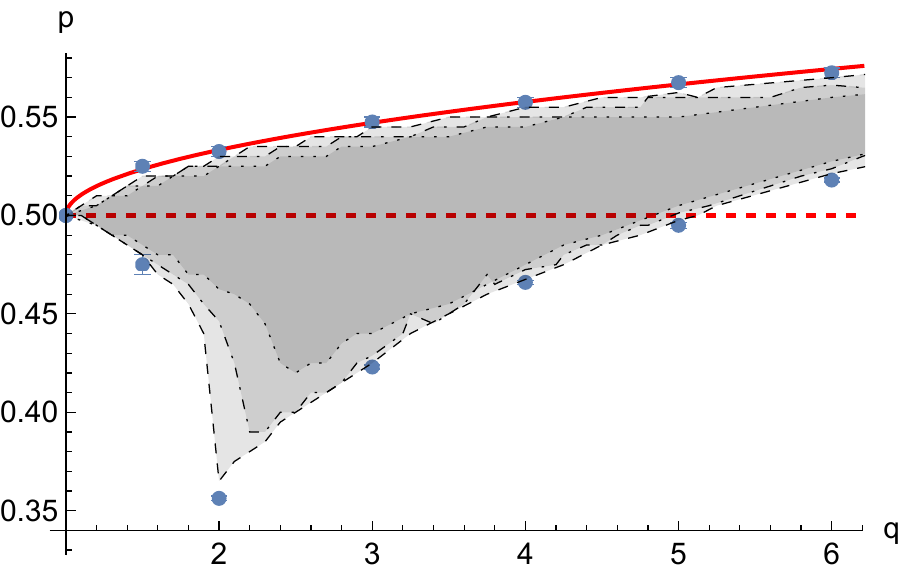}
  \caption{\label{fig12}(Colour on-line) $d=3$, $q>1$. Support $S_L$
    for $L=6$ (innermost region, dotted boundary), $L=8$ (dot-dashed
    boundary) and $S_{12}$ (dashed boundary). Red dashed line is
    $p=1/2$. Solid red curve is $p=0.50+0.0333\sqrt{q-1}$. Points
    indicate boundary of $S$.}
\end{figure}

For 3-dimensional systems the chaotic phase for $q>1$ is interesting
as well. In Fig.~\ref{fig12} we show the support sets for $L=6,8,12$
together with some estimated boundary points of the chaotic
phase. Again we find a very good approximation for the upper boundary,
$\bar S(q)=0.50+0.0333(1)\sqrt{q-1}$, for $q\lesssim 20$ (!).  For
$q\gtrsim 20$ the formula does not give a good prediction of the upper
bound and we do not know the shape of $S$ beyond this point.

\begin{figure}
  \includegraphics[width=0.483\textwidth]{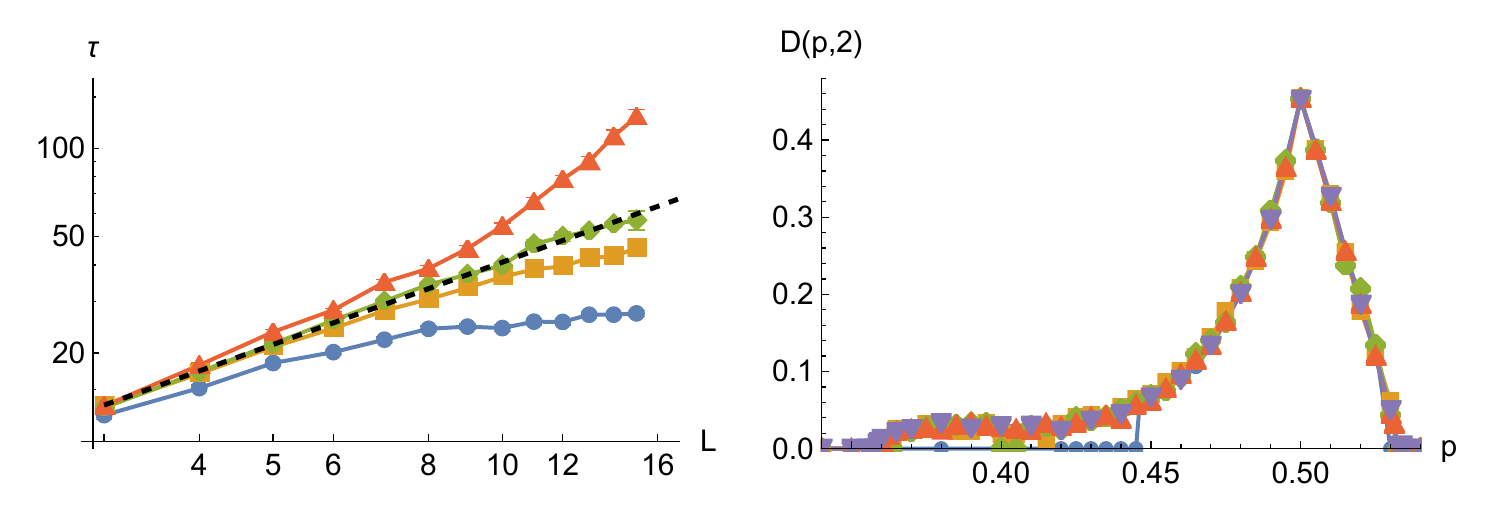}
  \caption{\label{fig13}(Colour on-line) $d=3$, $q=2$. Left: median
    coupling time $\tau$ versus $L$ for (upwards) $p=0.35$ (blue
    points), $p=0.355$ (orange squares), $p=0.35625$ (green diamonds)
    and $p=0.358091$ (red triangles). Fitted line (dashed) corresponds
    to $\tau = 4.8L^{0.93}$. Right: Damage $D(p,2)$ for
    $L=8,10,12,16,20$.  $D\to 0$ at $p\le 0.44$ for $L=8$ and at
    $p=0.40$ for $L=12$.}
\end{figure}

The lower boundary points $\underbar S(q)$ agree very well with the
critical inverse temperature of the $q$-state Potts model, after using
the relation $p_c(q)=1-e^{-\beta_c(q)}$.  Let us begin with $q=2$,
where $p_c(2)\approx 0.358091$. In Fig.~\ref{fig13} we show $\tau$
versus $L$ around this value and the damage function along $q=2$. In
this case it appears that $\tau$ grows exponentially at $p=0.358091$,
approximately $\tau = 10.3(3)\exp(0.168(4)L)$ for $L\ge 5$.
Decreasing $p$ only slightly, at $p=0.35625$, we find instead $\tau =
4.8(2)L^{0.93(2)}$ for $L\ge 3$, while at $p=0.355$ we estimate $\tau
= -14(1) + 22(1) \log L$. This suggests that $\underbar S(2)<p_c(2)$
though the difference is only about $0.5\%$.

\begin{figure}
  \includegraphics[width=0.483\textwidth]{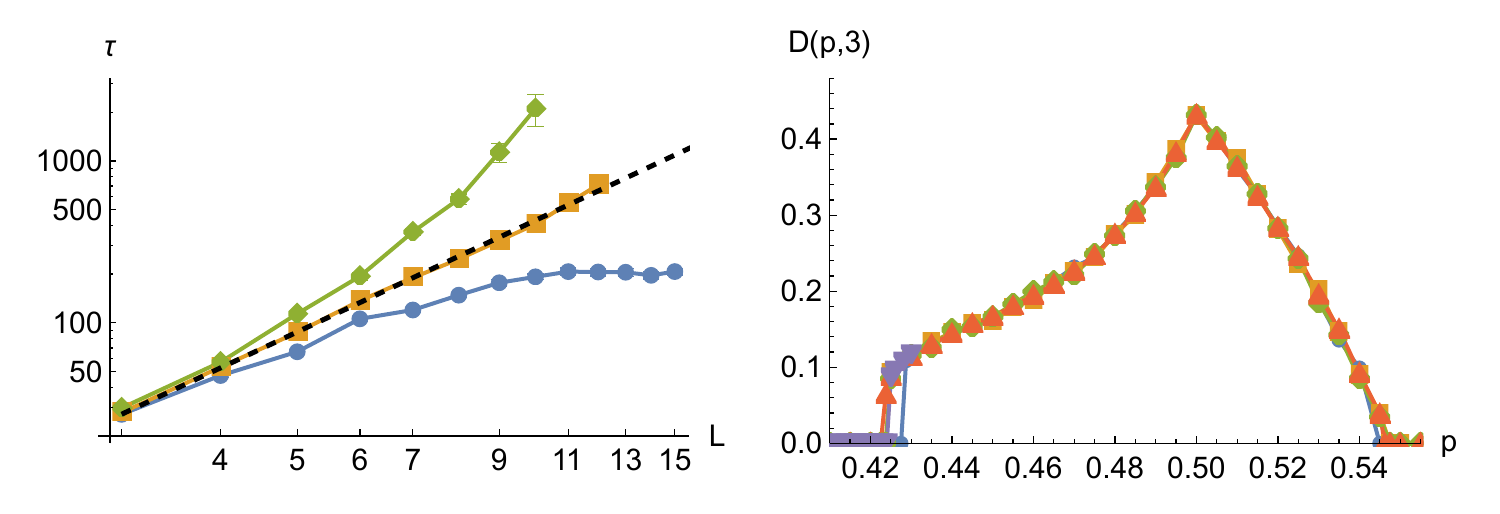}
  \caption{\label{fig14}(Colour on-line) $d=3$, $q=3$. Left: median
    coupling time $\tau$ versus $L$ for (upwards) $p=0.422$ (blue
    points), $p=0.4233$ (orange squares), $p=0.425$ (green
    diamonds). Fitted line (dashed) corresponds to $\tau =
    2.2L^{2.3}$. Right: Damage $D(p,3)$ for $L=8,10,12,16,20$.}
\end{figure}

In somewhat less detail, let us also consider $q=3$, recall that
$p_c(3)=0.42330(5)$. In Fig.~\ref{fig14} we again show $\tau$ near
$p_c(3)$ and the damage $D(p,3)$. At $p=0.4233$ we indeed see a
polynomial growth in the coupling time ($\tau = 2.2(3) L^{2.3(1)}$), a
clear exponential behaviour at $p=0.425$ ($\tau \propto \exp(0.6L)$),
and $\tau\lesssim 200$ at $p=0.422$. Hence $\underbar S(3)$ and
$p_c(3)$ may be equal but any difference is within $0.3\%$.

\begin{figure}
  \includegraphics[width=0.483\textwidth]{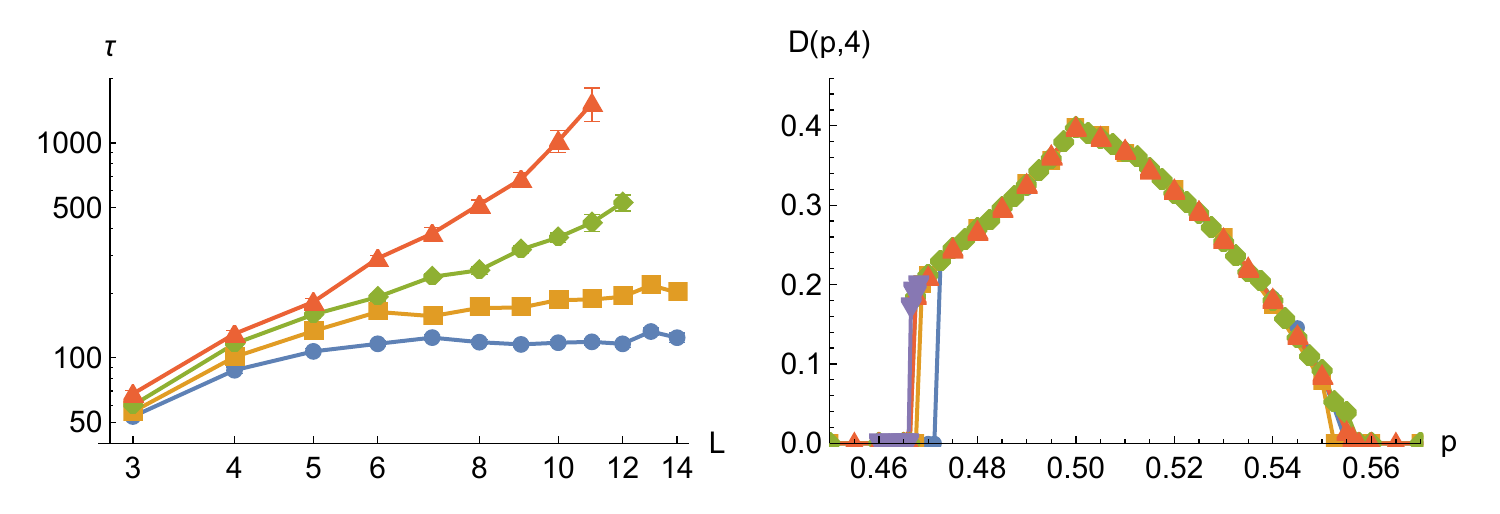}
  \caption{\label{fig15}(Colour on-line) $d=3$, $q=4$. Left: median
    coupling time $\tau$ versus $L$ for (upwards) $p=0.464$ (blue
    points), $p=0.465$ (orange squares), $p=0.466$ (green diamonds)
    and $p=0.467$ (red triangles). Right: Distance $D(p,4)$ for
    $L=8,10,12,16,20$. }
\end{figure}

For $q=4$, as shown in Fig.~\ref{fig15}, the finite-size effects in
$\tau$ complicate the picture and it is difficult to estimate
$\underbar S(4)$ with good accuracy. At $p=0.467$, just above the
estimate $p_c(4)=0.4666(3)$ we see a distinctly exponential growth
rate in $\tau$. At $p=0.465$, just below the estimate, $\tau$ actually
appears to be bounded. We show also $\tau$ for $p=0.466$ but it would
take larger $L$ to decide if it has polynomial growth. This gives us
the estimate $\underbar S(4) = 0.466(1)$. This estimate of $\underbar
S(4)$ then overlaps the estimate of $p_c(4)$.

\begin{figure}
  \includegraphics[width=0.483\textwidth]{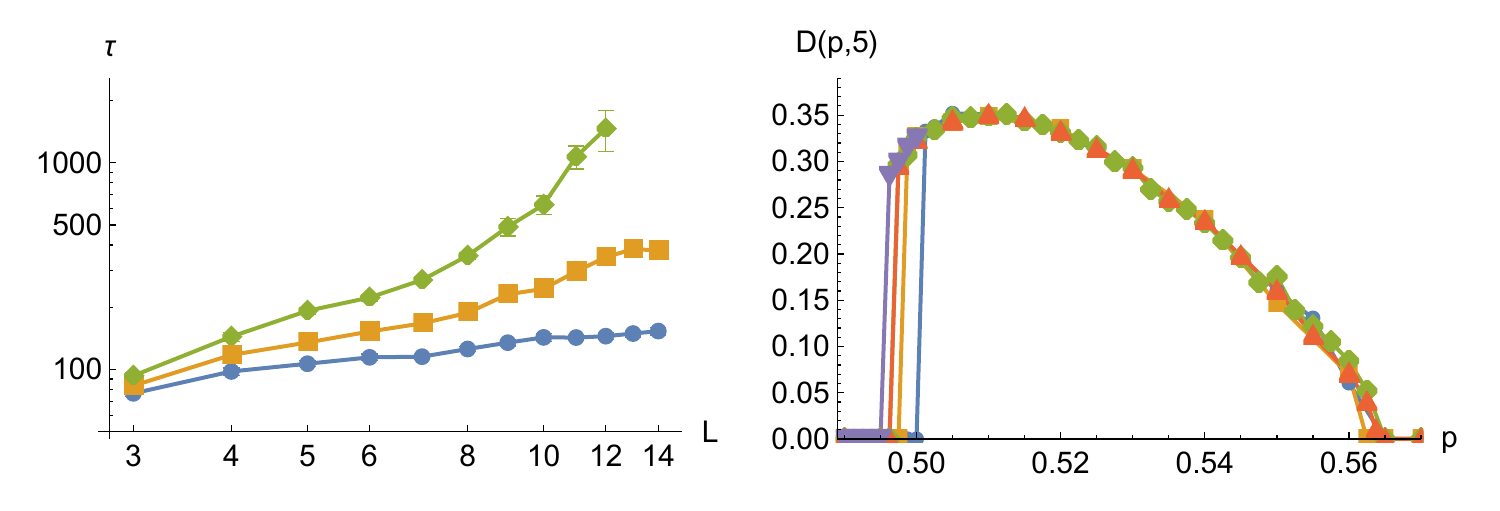}
  \caption{\label{fig16}(Colour on-line) $d=3$, $q=5$. Left: median
    coupling time $\tau$ versus $L$ for (upwards) $p=0.49375$ (blue
    points), $p=0.495$ (orange squares) and $p=0.49625$ (green
    diamonds). Right: Distance $D(p,5)$ for $L=8,10,12,16,20$. }
\end{figure}

We continue with $q=5$ and give an estimate of $\underbar S(5)$.  In
the left panel of Fig.~\ref{fig16} we see that $p=0.49625$ gives
distinct exponential growth in $\tau$ and at $p=0.49375$ the growth is
logarithmic. We also show $\tau$ for $p=0.495$ but the small-$L$
effects make it difficult to say which category it belongs to. We
obtain the estimate $\underbar S(5)=0.4950(13)$ which translates to
$\beta_c(5) = 0.6832(26)$, if we assume that $p_c$ and the lower bound
are equal. Analogously, we find for $q=6$ that the lower bound of the
chaotic phase is $\underbar S(6) = 0.518(1)$ suggesting
$\beta_c(6)=0.730(1)$ for the 6-state Potts model.

In the right panel of Fig.~\ref{fig16}, showing $D(p,5)$, we note that
the maximum is no longer located at $p=1/2$, but rather at $p\approx
0.51$.  In fact, our data suggest that the maximum is located at
$p=1/2$ for $q\le 4.5$ but at some $p$ larger than $1/2$ for $q> 4.5$.

\begin{figure}
  \includegraphics[width=0.483\textwidth]{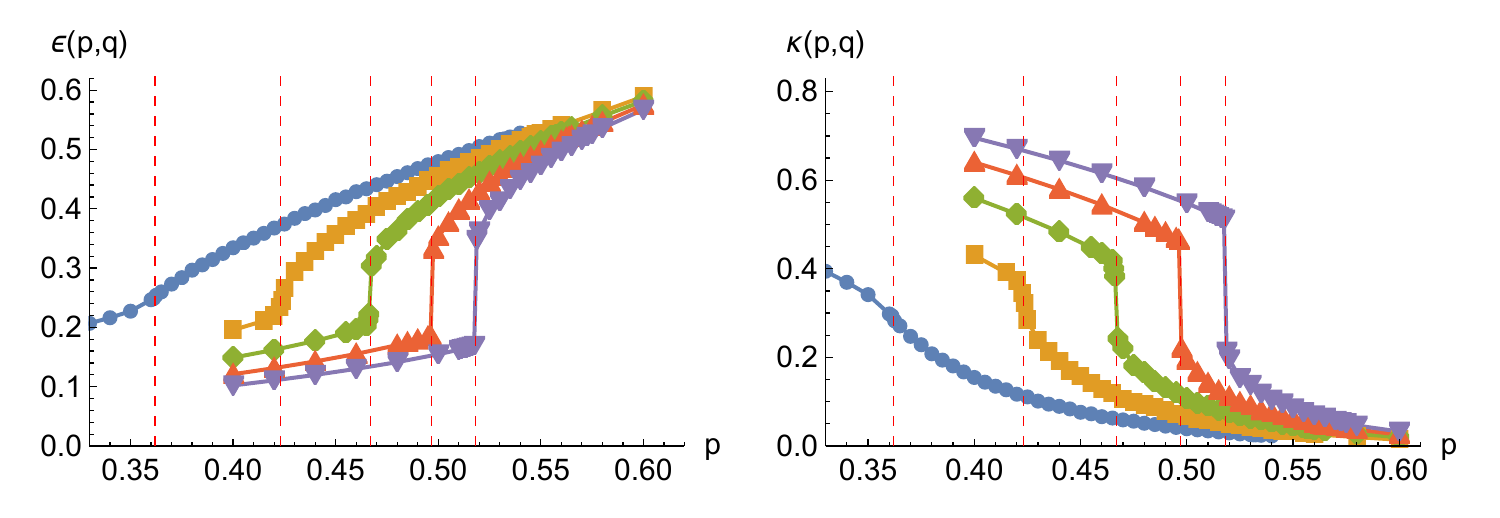}
  \caption{\label{fig17}(Colour on-line) $d=3$, $L=16$.  Normalised
    mean number of edges $\epsilon(p,q)$ (left panel) and components
    $\kappa(p,q)$ (right panel) versus $p$ for $q=2,3,4,5,6$
    (rightwards in plots).  Dashed vertical lines at $\underbar
    S_{16}(q)$.}
\end{figure}

The phenomenon of an emerging discontinuity in $D(p,q)$ is also
reflected in other fundamental quantities of the random-cluster model.
In Fig.~\ref{fig17} we show $\epsilon(p,q)$ and $\kappa(p,q)$ versus
$p$ for $q=2,3,4,5,6$ and $L=16$. A discontinuity is clearly emerging
with increasing $q$, becoming more distinct at $q=4$, and it is
located at exactly the same point where the damage drops to zero.
Thus it is perhaps not surprising that the critical temperature of the
$q$-Potts model is connected to this point. On the other hand, the
upper bound of $S$ does not seem to have a corresponding effect on
$\epsilon$ and $\kappa$. For rigorous results on the discontinuity of
$\epsilon$, see Section 7.5 of Ref.~\cite{grimmett:06})

\section{Conclusions}
We have investigated the phenomenon of damage spreading in the
random-cluster model for grids of dimensions two and three with
periodic boundary conditions. The chaotic phase, where the coupling
time $\tau$ grows exponentially with $L$, appears as an island in the
$p,q$-plane and here the damage function takes positive values and its
average remains stable for many time steps.

For 2-dimensional grids the damage function has a global cusp-shaped
maximum at exactly $p=p_c(q)=\sqrt{q}/(1+\sqrt{q})$. When $q\lesssim
0.75$ there are also local maxima, again cusp-shaped, at $p=q/(1+q)$
and $p=1/2$. For comparison, we observe that the edge-flip
probability, under Metropolis updating, also has local cusp-like
maxima at $p=q/(1+q)$ and $p=1/2$.  We have estimated lower and upper
bounds for the chaotic phase and suggested an approximation formula for
$\bar S(q)$ when $q<1$.

For 3-dimensional grids the damage function has a richer behaviour,
depending on $q$. Notably, for $q\lesssim 0.46$ the damage function
has a cusp-shaped local maximum at $p=q/(1+q)$ and we observe a cusp
also at $p=\sqrt{q}/(1+\sqrt{q})$. The chaotic phase has a more
complex shape for $q<1$ but we give good approximation formulas for
the upper bound when $q<20$ and for the lower bound when $q<0.75$.
When $q>2$ is integer, the lower bound appears remarkably close to the
critical point $p_c(q)$ of the $q$-state Potts model. They differ
slightly for $q=2$ but estimates of $p_c(q)$ and $\underbar S(q)$
overlap for $q=3,4$. For $q>2$ the damage function (with fixed $q$)
develops a discontinuity at $\underbar S(q)$ and the same effect is
observed for both $\epsilon$ and $\kappa$ at this point.



\begin{thebibliography}{20}
\expandafter\ifx\csname natexlab\endcsname\relax\def\natexlab#1{#1}\fi
\expandafter\ifx\csname bibnamefont\endcsname\relax
  \def\bibnamefont#1{#1}\fi
\expandafter\ifx\csname bibfnamefont\endcsname\relax
  \def\bibfnamefont#1{#1}\fi
\expandafter\ifx\csname citenamefont\endcsname\relax
  \def\citenamefont#1{#1}\fi
\expandafter\ifx\csname url\endcsname\relax
  \def\url#1{\texttt{#1}}\fi
\expandafter\ifx\csname urlprefix\endcsname\relax\def\urlprefix{URL }\fi
\providecommand{\bibinfo}[2]{#2}
\providecommand{\eprint}[2][]{\url{#2}}

\bibitem[{\citenamefont{Derrida and Pomeau}(1986)}]{derrida:86}
\bibinfo{author}{\bibfnamefont{B.}~\bibnamefont{Derrida}} \bibnamefont{and}
  \bibinfo{author}{\bibfnamefont{Y.}~\bibnamefont{Pomeau}},
  \bibinfo{journal}{Europhys. Lett.} \textbf{\bibinfo{volume}{1}},
  \bibinfo{pages}{45} (\bibinfo{year}{1986}).

\bibitem[{\citenamefont{Stanley et~al.}(1987)\citenamefont{Stanley, Stauffer,
  Kert\'esz, and Herrmann}}]{stanley:87}
\bibinfo{author}{\bibfnamefont{H.~E.} \bibnamefont{Stanley}},
  \bibinfo{author}{\bibfnamefont{D.}~\bibnamefont{Stauffer}},
  \bibinfo{author}{\bibfnamefont{J.}~\bibnamefont{Kert\'esz}},
  \bibnamefont{and} \bibinfo{author}{\bibfnamefont{H.~J.}
  \bibnamefont{Herrmann}}, \bibinfo{journal}{Phys. Rev. Lett.}
  \textbf{\bibinfo{volume}{59}}, \bibinfo{pages}{2326} (\bibinfo{year}{1987}).

\bibitem[{\citenamefont{Derrida and Weisbuch}(1987)}]{derrida:87}
\bibinfo{author}{\bibfnamefont{B.}~\bibnamefont{Derrida}} \bibnamefont{and}
  \bibinfo{author}{\bibfnamefont{G.}~\bibnamefont{Weisbuch}},
  \bibinfo{journal}{Europhys. Lett.} \textbf{\bibinfo{volume}{4}},
  \bibinfo{pages}{657} (\bibinfo{year}{1987}).

\bibitem[{\citenamefont{de~Arcangelis et~al.}(1989)\citenamefont{de~Arcangelis,
  Coniglio, and Herrmann}}]{arcangelis:89}
\bibinfo{author}{\bibfnamefont{L.}~\bibnamefont{de~Arcangelis}},
  \bibinfo{author}{\bibfnamefont{A.}~\bibnamefont{Coniglio}}, \bibnamefont{and}
  \bibinfo{author}{\bibfnamefont{H.~J.} \bibnamefont{Herrmann}},
  \bibinfo{journal}{Europhys. Lett.} \textbf{\bibinfo{volume}{9}},
  \bibinfo{pages}{749} (\bibinfo{year}{1989}).

\bibitem[{\citenamefont{Campbell and de~Arcangelis}(1991)}]{campbell:91}
\bibinfo{author}{\bibfnamefont{I.~A.} \bibnamefont{Campbell}} \bibnamefont{and}
  \bibinfo{author}{\bibfnamefont{L.}~\bibnamefont{de~Arcangelis}},
  \bibinfo{journal}{Physica A} \textbf{\bibinfo{volume}{178}},
  \bibinfo{pages}{29} (\bibinfo{year}{1991}).

\bibitem[{\citenamefont{Campbell}(1993)}]{campbell:93}
\bibinfo{author}{\bibfnamefont{I.~A.} \bibnamefont{Campbell}},
  \bibinfo{journal}{Europhys. Lett.} \textbf{\bibinfo{volume}{21}},
  \bibinfo{pages}{959} (\bibinfo{year}{1993}).

\bibitem[{\citenamefont{Lundow and Campbell}(2012)}]{lundow:12}
\bibinfo{author}{\bibfnamefont{P.~H.} \bibnamefont{Lundow}} \bibnamefont{and}
  \bibinfo{author}{\bibfnamefont{I.~A.} \bibnamefont{Campbell}},
  \bibinfo{journal}{Phys. Rev. E} \textbf{\bibinfo{volume}{86}},
  \bibinfo{pages}{041121} (\bibinfo{year}{2012}).

\bibitem[{\citenamefont{Bernard et~al.}(2010)\citenamefont{Bernard, Chanal, and
  Krauth}}]{bernard:10}
\bibinfo{author}{\bibfnamefont{E.~P.} \bibnamefont{Bernard}},
  \bibinfo{author}{\bibfnamefont{C.}~\bibnamefont{Chanal}}, \bibnamefont{and}
  \bibinfo{author}{\bibfnamefont{W.}~\bibnamefont{Krauth}},
  \bibinfo{journal}{Europhys. Lett.} \textbf{\bibinfo{volume}{92}},
  \bibinfo{pages}{60004} (\bibinfo{year}{2010}).

\bibitem[{\citenamefont{Puzzo et~al.}(2013)\citenamefont{Puzzo, Saracco, and
  Albano}}]{rubio:13}
\bibinfo{author}{\bibfnamefont{M.~L.~R.} \bibnamefont{Puzzo}},
  \bibinfo{author}{\bibfnamefont{G.~P.} \bibnamefont{Saracco}},
  \bibnamefont{and} \bibinfo{author}{\bibfnamefont{E.~V.}
  \bibnamefont{Albano}}, \bibinfo{journal}{Physica A}
  \textbf{\bibinfo{volume}{392}}, \bibinfo{pages}{2680} (\bibinfo{year}{2013}).

\bibitem[{\citenamefont{Puzzo et~al.}(2016)\citenamefont{Puzzo, Saracco, and
  Bab}}]{rubio:16}
\bibinfo{author}{\bibfnamefont{M.~L.~R.} \bibnamefont{Puzzo}},
  \bibinfo{author}{\bibfnamefont{G.~P.} \bibnamefont{Saracco}},
  \bibnamefont{and} \bibinfo{author}{\bibfnamefont{M.~A.} \bibnamefont{Bab}},
  \bibinfo{journal}{Physica A} \textbf{\bibinfo{volume}{444}},
  \bibinfo{pages}{476} (\bibinfo{year}{2016}).

\bibitem[{\citenamefont{Fortuin and Kasteleyn}(1972)}]{FK:72}
\bibinfo{author}{\bibfnamefont{C.~M.} \bibnamefont{Fortuin}} \bibnamefont{and}
  \bibinfo{author}{\bibfnamefont{P.~W.} \bibnamefont{Kasteleyn}},
  \bibinfo{journal}{Physica} \textbf{\bibinfo{volume}{57}},
  \bibinfo{pages}{536} (\bibinfo{year}{1972}).

\bibitem[{\citenamefont{Grimmett}(2006)}]{grimmett:06}
\bibinfo{author}{\bibfnamefont{G.}~\bibnamefont{Grimmett}},
  \emph{\bibinfo{title}{The random-cluster model}}
  (\bibinfo{publisher}{Springer}, \bibinfo{year}{2006}).

\bibitem[{\citenamefont{M\"uller-Krumbhaar and Binder}(1973)}]{binder:73}
\bibinfo{author}{\bibfnamefont{H.}~\bibnamefont{M\"uller-Krumbhaar}}
  \bibnamefont{and} \bibinfo{author}{\bibfnamefont{K.}~\bibnamefont{Binder}},
  \bibinfo{journal}{J. Stat. Phys.} \textbf{\bibinfo{volume}{8}}
  (\bibinfo{year}{1973}).

\bibitem[{\citenamefont{Beffara and Duminil-Copin}(2011)}]{beffara:12}
\bibinfo{author}{\bibfnamefont{V.}~\bibnamefont{Beffara}} \bibnamefont{and}
  \bibinfo{author}{\bibfnamefont{H.}~\bibnamefont{Duminil-Copin}},
  \bibinfo{journal}{Probab. Theory Relat. Fields}
  \textbf{\bibinfo{volume}{153}}, \bibinfo{pages}{511} (\bibinfo{year}{2011}).

\bibitem[{\citenamefont{Qian et~al.}(2005)\citenamefont{Qian, Deng, and
  Bl\"ote}}]{blote:05}
\bibinfo{author}{\bibfnamefont{X.}~\bibnamefont{Qian}},
  \bibinfo{author}{\bibfnamefont{Y.}~\bibnamefont{Deng}}, \bibnamefont{and}
  \bibinfo{author}{\bibfnamefont{H.~W.~J.} \bibnamefont{Bl\"ote}},
  \bibinfo{journal}{Phys. Rev. E} \textbf{\bibinfo{volume}{71}},
  \bibinfo{pages}{016709} (\bibinfo{year}{2005}).

\bibitem[{\citenamefont{Ferrenberg et~al.}(2018)\citenamefont{Ferrenberg, Xu,
  and Landau}}]{ferrenberg:18}
\bibinfo{author}{\bibfnamefont{A.~M.} \bibnamefont{Ferrenberg}},
  \bibinfo{author}{\bibfnamefont{J.}~\bibnamefont{Xu}}, \bibnamefont{and}
  \bibinfo{author}{\bibfnamefont{D.~P.} \bibnamefont{Landau}},
  \bibinfo{journal}{Phys. Rev. E} \textbf{\bibinfo{volume}{97}},
  \bibinfo{pages}{043301} (\bibinfo{year}{2018}).

\bibitem[{\citenamefont{Lundow and Markstr\"om}(2009)}]{lundow:09}
\bibinfo{author}{\bibfnamefont{P.~H.} \bibnamefont{Lundow}} \bibnamefont{and}
  \bibinfo{author}{\bibfnamefont{K.}~\bibnamefont{Markstr\"om}},
  \bibinfo{journal}{Open Phys.} \textbf{\bibinfo{volume}{7}},
  \bibinfo{pages}{490} (\bibinfo{year}{2009}).

\bibitem[{\citenamefont{Janke and Villanova}(1997)}]{janke:97}
\bibinfo{author}{\bibfnamefont{W.}~\bibnamefont{Janke}} \bibnamefont{and}
  \bibinfo{author}{\bibfnamefont{R.}~\bibnamefont{Villanova}},
  \bibinfo{journal}{Nucl. Phys. B} \textbf{\bibinfo{volume}{489}},
  \bibinfo{pages}{679} (\bibinfo{year}{1997}).

\bibitem[{\citenamefont{Chatelain et~al.}(2005)\citenamefont{Chatelain, Berche,
  Janke, and Berge}}]{chatelain:05}
\bibinfo{author}{\bibfnamefont{C.}~\bibnamefont{Chatelain}},
  \bibinfo{author}{\bibfnamefont{B.}~\bibnamefont{Berche}},
  \bibinfo{author}{\bibfnamefont{W.}~\bibnamefont{Janke}}, \bibnamefont{and}
  \bibinfo{author}{\bibfnamefont{P.-E.} \bibnamefont{Berge}},
  \bibinfo{journal}{Nucl. Phys. B} \textbf{\bibinfo{volume}{719}},
  \bibinfo{pages}{275} (\bibinfo{year}{2005}).

\bibitem[{\citenamefont{Babaev and Murtazaev}(2015)}]{babaev:15}
\bibinfo{author}{\bibfnamefont{A.~B.} \bibnamefont{Babaev}} \bibnamefont{and}
  \bibinfo{author}{\bibfnamefont{A.~K.} \bibnamefont{Murtazaev}},
  \bibinfo{journal}{Low Temp. Phys.} \textbf{\bibinfo{volume}{41}},
  \bibinfo{pages}{608} (\bibinfo{year}{2015}).

\end{thebibliography}

\end{document}